\documentclass[1p]{elsarticle}

\usepackage{lineno,hyperref}
\journal{PRL}

\usepackage{graphicx}
\usepackage{epstopdf}
\usepackage{textcomp}
\usepackage[utf8]{inputenc}
\usepackage[usenames, dvipsnames]{color}
\usepackage[normalem]{ulem}

\begin{document}

\begin{frontmatter}

%\preprint{}

\title{Shear softening in a metallic glass: first principle local stress analysis}

\author{I.~Lobzenko$^a$, Y.~Shiihara$^a$, T.~Iwashita$^b$, T.~Egami$^{c,d}$}

\address{$^a$Toyota Technological Institute, Hisakata, Tempaku-ku, Nagoya 468-8511, Japan}
\address{$^b$ Oita University, Dannoharu, Oita 870-1192, Japan}
\address{$^c$ University of Tennessee, Knoxville, Tennessee 37996, USA}
\address{$^d$ Oak Ridge National Laboratory, Oak Ridge, TN 37831, USA}

%%\date{\today}

\begin{abstract}
  Metallic glasses deform elastically under stress. However, the atomic-level origin of elastic properties of metallic glasses remain unclear. In this paper using {\em ab initio} molecular dynamics simulations of the Cu$_{50}$Zr$_{50}$ metallic glass under shear strain, we show that the heterogeneous stress relaxation results in the increased charge transfer from Zr to Cu atoms, enhancing the softening of the shear modulus. Changes in compositional short-range order and atomic position shifts due to the non-affine deformation are discussed. It is shown that the Zr subsystem exhibits a stiff behavior, whereas the displacements of Cu atoms from their initial positions, induced by the strain, provide the stress drop and softening.
\end{abstract}

% \begin{keyword}
%   Metallic glass\sep First principle calculations\sep Elastic properties\sep Atomic stress\sep Compositional short-range order
% \end{keyword}

\end{frontmatter}

%\maketitle

%%\newpage
%%\linenumbers

%%\linenumbers

%%\section{Introduction}

% {\tHGHL ... }

%\begin{linenumbers}

Amorphous materials formed by metal atoms, usually referred to as metallic glasses (MG) or glassy metals, have gained significant attention after the procedure of mass production of the bulk form of such materials had been developed in the 1990s~\cite{inoue1990,inoue2011}, made possible by the discovery of glass-forming alloys requiring relatively low cooling rates of $<$100~K$\cdot$s$^{-1}$~\cite{johnson1999}. This enabled broad applications of MGs, for instance, for nanoimprinted technology, bio-implants, and coating, to name a few examples~\cite{inoue2011,kumar2011}.

Along with the applications, metallic glasses have received intense scientific interest. One of the striking features of MGs is shear modulus softening (SMS), namely the shear modulus of MGs is significantly lower than that of crystalline counterparts~\cite{davis1976}. This phenomenon has been extensively studied
theoretically with assumptions being supported by computer simulations in a number of early~\cite{weaire1971,suzuki1985} and most recent~\cite{dmowski2010,egamiReview2013,sawPRL2016,yuJPCL2016} works, and the microscopic origin has been attributed to non-affine heterogeneous atomic motions during deformation.
However, the precise atomic mechanisms of SMS remain elusive. Particularly little quantum level understanding of SMS has been achieved~\cite{egamiReview2013}.

In the case of crystalline materials, inelastic behavior is governed by the well-defined lattice defects in the periodic structure. However, attempts of defining defects in glasses face a formidable conceptual and practical barrier because of structural disorder. A number of theories of deformation in MG were proposed, among which the most widely used relies on the so-called shear transformation zone (STZ)~\cite{argon1979_STZ1,argon1979_STZ2,langerSTZ2004}.
It attributes the deformation in amorphous material to the emergence and development of regions with higher mobility. Despite advances in the STZ theory, a number of its key features remain veiled. For example, many simulations and experiments show contradictory results for STZ sizes, from a few atoms to several hundred, and also the precise physical picture of the STZ is missing~\cite{iwashitaNatComm_Ref18,iwashitaNatComm_Ref19,iwashitaNatComm_Ref20, iwashitaNatComm_Ref21}. Furthermore, the particular mechanism of STZ emergence is yet to be understood.

Advances in modern computational methods allow one to obtain insights into the deformation behavior of individual atoms. One of the most accurate atomistic modeling methods is based on the density functional theory (DFT), which provides the evaluation of structural parameters with accuracy up to 1\%~\cite{kohnNobelLecture1999}.
This approach is effective in defining the starting point of shear transformation zone in metallic glass since relatively small structures (order of 20 atoms) are involved in the STZ~\cite{fanIwashita2015}. In order to examine the local stress state around the STZ, which is highly heterogeneous at the microscopic scale, we use the atomic-level stresses analysis~\cite{egamiConn1980,egamiConn2014,egamiConn2017}, coupled with the DFT calculation~\cite{nicholsonFPStress2013}.

In the current work, we study Cu$_{50}$Zr$_{50}$ glass under shear strain. The choice of the system is dictated by the fact that a large amount of data are available for CuZr alloys in both crystalline and glassy structures~\cite{nicholsonFPStress2013,matternCuZr2008_1}. It is known that the CuZr glass could be obtained in a wide composition range. Therefore, the influence of the stoichiometry presented in the current work may be verified by future experimental work.

This paper is organized as follows. At first, the details on the structures under study and settings for the first principle calculations are given. Next, the main results are presented and discussed with emphasis on the fundamental properties of the structure leading to shear modulus softening.

\begin{figure*}
  \centering
  \includegraphics[width=0.95\textwidth]{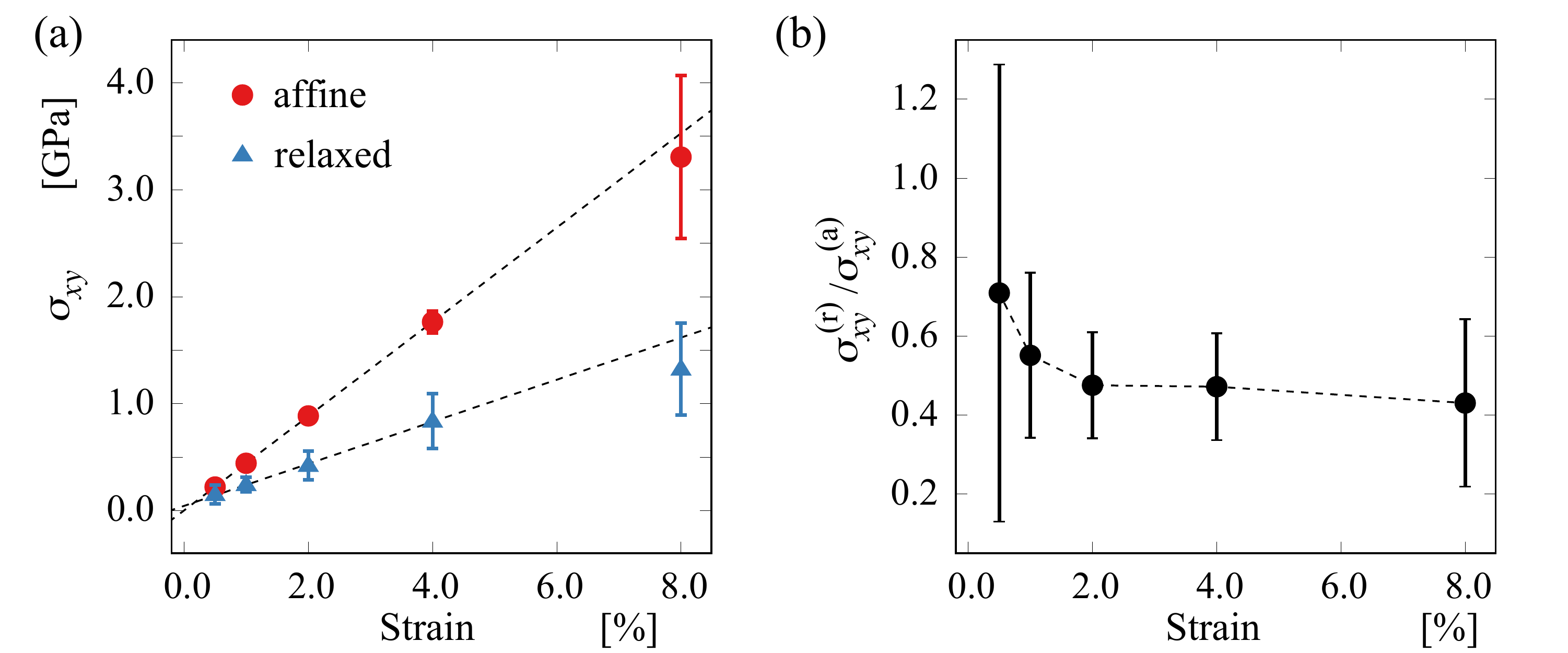}
  \caption{(a) Stress - strain curves of Cu$_{50}$Zr$_{50}$ glass for affine and relaxed structures. (b) Ratio of relaxed stress $\sigma_{xy}^\mathrm{(r)}$ to affine stress $\sigma_{xy}^\mathrm{(a)}$ for various shear strains. Vertical bars show errors due to averaging (see discussion in text).}
  \label{g-StrStr}
\end{figure*}

%[hb]
\begin{figure*}
  \centering
  \includegraphics[width=0.95\textwidth]{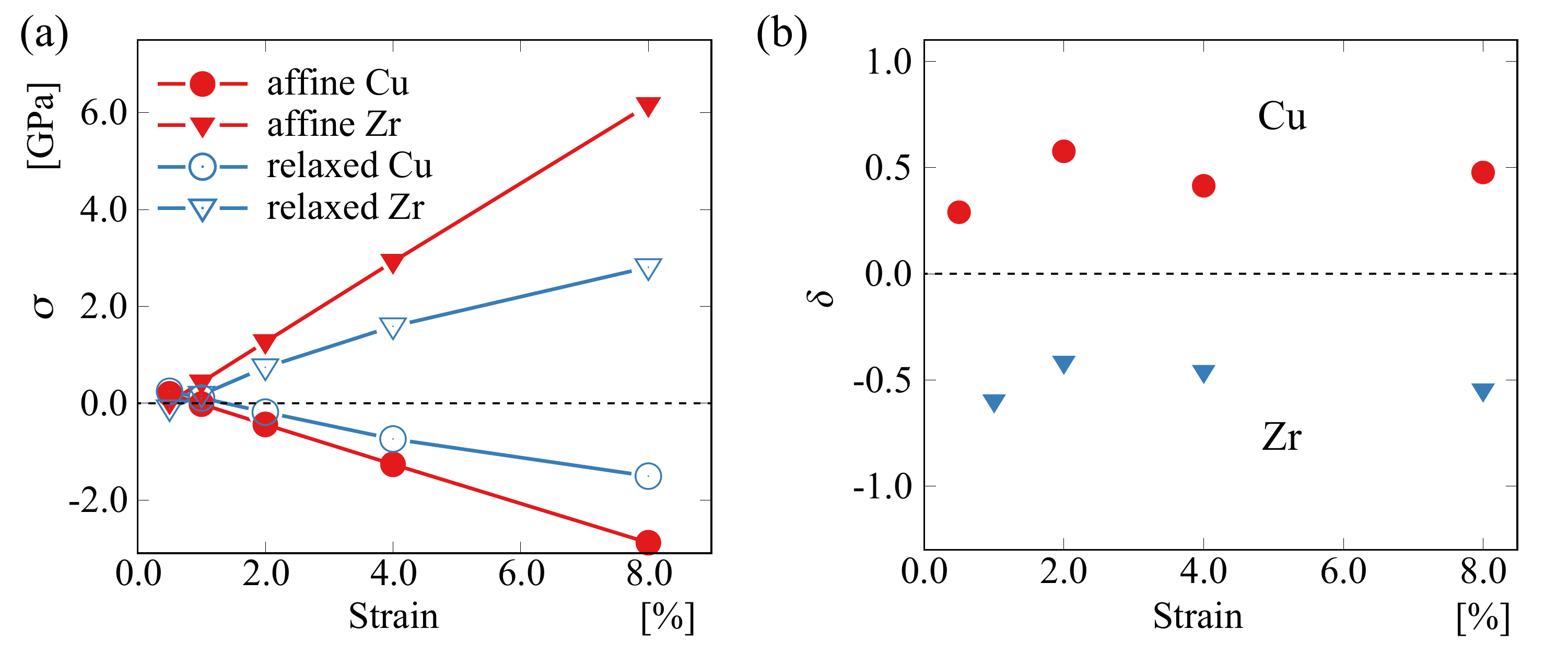}
  \caption{{\em(Color online)} (a) Shear stress in the affine and relaxed structures for Cu$_{50}$Zr$_{50}$ under shear strain. Red filled circles (triangles) show Cu (Zr) contribution to the stress in the affine structure, while blue empty circles (triangles) show Cu (Zr) contribution to stress in the relaxed structure. (b) Deviation ($\delta=(\sigma_{\mathit{relaxed}}-\sigma_\mathit{affine})/|\sigma_\mathit{affine}|$) of the  $xy$ component of stress in the relaxed system from the $xy$ component of stress in the affine system. Red circles (blue triangles) represent Cu (Zr) atoms. }
  \label{g-dStr}
\end{figure*}

% \section{Simulation details}
%
% \subsection{Equilibrium structure preparation} \label{ssec:strPrep}

Initially, a random atomic configuration of 96 atoms was prepared at a density of
57.1~nm$^{-3}$~\cite{matternCuZr2008}. The first principle molecular dynamics (FPMD) simulation at 3000~K for 2~ps was performed under the NVE ensemble with periodic boundary conditions to thermalize the original structure, thus obtaining an equilibrium liquid structure. The time step of the simulations was 2~fs. The system was quenched to a supercooled state at 1200~K, which led to the relatively stable glass structure, then FPMD was performed for 1~ps. Next the system was gradually cooled down to a glass state at 700~K with a cooling rate of 0.5~K/fs, and in addition, another thermalization at 300~K was performed for 1~ps to stabilize the glass structure. Finally, the glass structure was relaxed to 0 K by applying energy minimization using the conjugate gradient method under zero macroscopic stress, allowing the box size and shape to vary during the iterations. The described procedure provides us with an equilibrium structure of the metallic glass ready for subsequent analyses.

% \subsection{Basic {\em ab initio} setup}

In our study, the quantum mechanical approach in the framework of density functional theory was used, which is implemented in the Vienna Ab initio Simulation Package (VASP)~\cite{vasp}. The generalized gradient approximation was used for the exchange-correlation energy, which is essential for the achievement of high accuracy. The energy cutoff regulating the number of basis functions was set to 410~eV. Because of the relatively large size of the system, only the $\Gamma$-point was used in the reciprocal space. The convergence of the self-consistent calculation was enhanced by the
Methfessel-Paxton method~\cite{vasp_mp} with the smearing of $0.1$~eV. In the energy minimization process, structures were optimized until the atomic forces become less than $0.01$~eV/\AA.

% \subsection{Modeling of response to strain} \label{ssec:strainAver}

To model the response of the CuZr system to strain, the athermal quasi-static shear (AQS) simulations, which are usually used in the framework of classical molecular dynamics, was performed at the quantum mechanics level using the FPMD. Firstly the glass structure at 0~K obtained by cooling is uniformly deformed with a simple shear strain $\varepsilon_{xy}$ (affine deformation). Next, atomic positions optimization is performed with the box geometry fixed. That results in the stress relaxation from the affine state during the energy minimization. To improve statistics, four different original structures, each of which is deformed independently at the same strain in six different directions, were prepared and the results were averaged over 24 samples (=~$4\times6$) in total.

In the process of analyzing the mechanical properties of materials under strain, local stress calculation is a standard tool to unveil atomic-level correlations. Nevertheless, until recently that tool was available only in the classical approximation. Lately, several approaches were proposed allowing the calculation of atomic level stress in a quantum-mechanical framework (see~\cite{nicholsonFPStress2013} for the detailed discussion of differences between those approaches). In particular, in this technique, the local stress is obtained as the strain derivative of local energy assigned to a single atom, as it is done in \cite{shiiharaLS2010} and \cite{nicholsonFPStress2013}.
The calculations were performed using the Open source package for Material eXplorer (OpenMX), which utilizes the orbital-based energy decomposition scheme~\cite{openMX}.
Such approach allows direct calculation of the derivatives of atomic energy with respect to the strain tensor.

% \section{Results}
%
% \subsection{Stress - strain}

Figure~\ref{g-StrStr} (a) shows stress as a function of strain for several shear strains $(0.005, 0.01,$ $0.02, 0.04, 0.08)$. To capture the effect of shear modulus softening~\cite{egamiReview2013}, under particular strain we analyzed two structures:
the {\em affine} one (obtained directly from the original structure by applying affine strain) and
the {\em relaxed} one (obtained after performing the optimization of atomic positions in the deformed structure). We see that stress is linearly proportional to applied strain for both cases of deformation. The estimated shear modulus is $44$~GPa for the affine deformation and $21$~GPa for the relaxed deformation. That is to say, the stress decreases during the relaxation process from the affine state by about 50\%, as shown in Fig~\ref{g-StrStr}~(b), which is consistent with previous results based on classical AQS simulations (see~\cite{nakamura2014} as an example).

%% *******
%% ******* ORIGINALY FIG 1
%% *******

In order to study the origin of the stress drop, we decomposed the macroscopic stress into the contributions of individual atoms and investigated the responses of the atomic-level stress under the shear for Cu and Zr. As shown in Fig.~\ref{g-dStr}~(a), the local stress analysis revealed a fascinating behavior. The local stress of Zr increases positively, as expected, with the strain for affine and relaxed structures, whereas the local stress of Cu becomes more negative with the strain, which is a quite unusual opposite behavior. Also, it can be seen that the relaxation decreases the magnitude of the stress on each atom, and the deviation,
$\delta = (\sigma_{relaxed}-\sigma_{affine})/\left| \sigma_{affine} \right|$
of the stress from the affine stress is plotted in Fig.~\ref{g-dStr}~(b). Interestingly Zr atoms show a negative contribution to the stress drop, while Cu atoms have a positive contribution. The origin of this behavior will be discussed later. At the highest strain of 8\%, the change in the stress corresponds to 1.38~GPa for Cu atoms and -3.31~GPa for Zr atoms, thus leading to the stress drop of -1.93~GP in total. Therefore, we see that the change in the local environment around Zr atoms should be mainly contributing to the total stress drop.

%% *******
%% ******* ORIGINALY FIG 2
%% *******

% \subsection{Atomic-level stresses and charge transfer in deformation}

\begin{figure}
  \centering
  \includegraphics[width=0.55\textwidth]{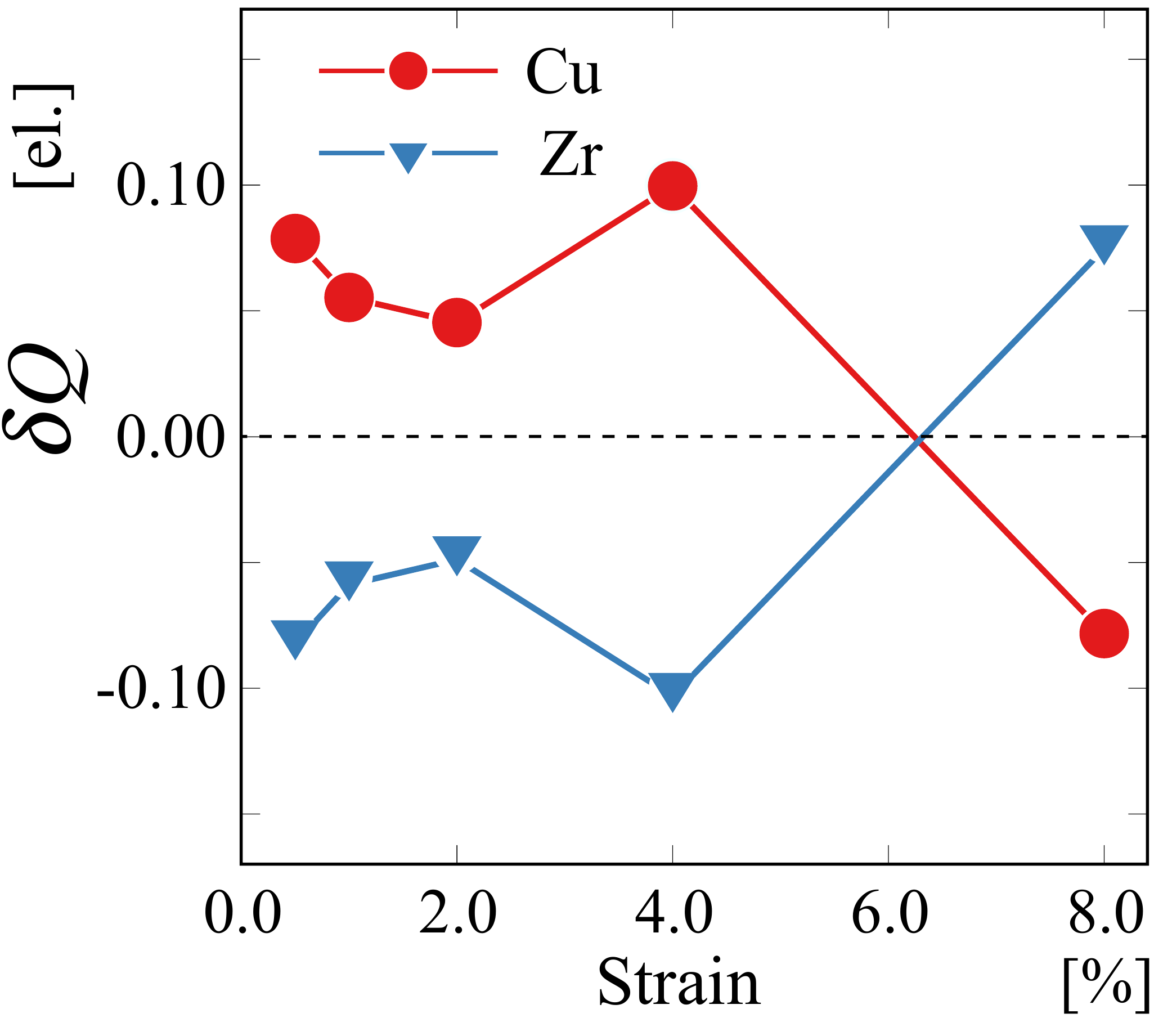}
  \caption{{\em(Color online)} Averaged total charge difference ($\delta Q$, relaxed to affine) expressed in number of electrons for Cu (red circles) and Zr (blue triangles) atoms with shear strain. }
  \label{g-dQonStrain}
\end{figure}

We investigate the role of charge on the stress drop under deformation. Figure~\ref{g-dQonStrain} shows the strain dependence on the charge transfer, $\delta Q$, between the affine and the relaxed states for Cu and Zr atoms. Units of $\delta Q$ are number of electrons per atom. At small strains ($\varepsilon \leq$ 4\%) Cu atoms gain more charge during relaxation, whereas Zr atoms lose it. These results indicate that the non-affine relaxation enhances charge transfer from Zr to Cu, which is explained by the fact that the original structure obtained by rapid quench is unstable and deformation leads to the formation of more bonds between Cu and Zr, resulting in a more stable structure. At the highest strain of 8\%, the sign of $\delta Q$ becomes opposite, which may be due to the plastic flow far from the elastic region which tends to increase the effective temperature of the system and rejuvenate it~\cite{dmowskiActaMater2010}.

\begin{figure*}
  \centering
  \includegraphics[width=0.55\textwidth]{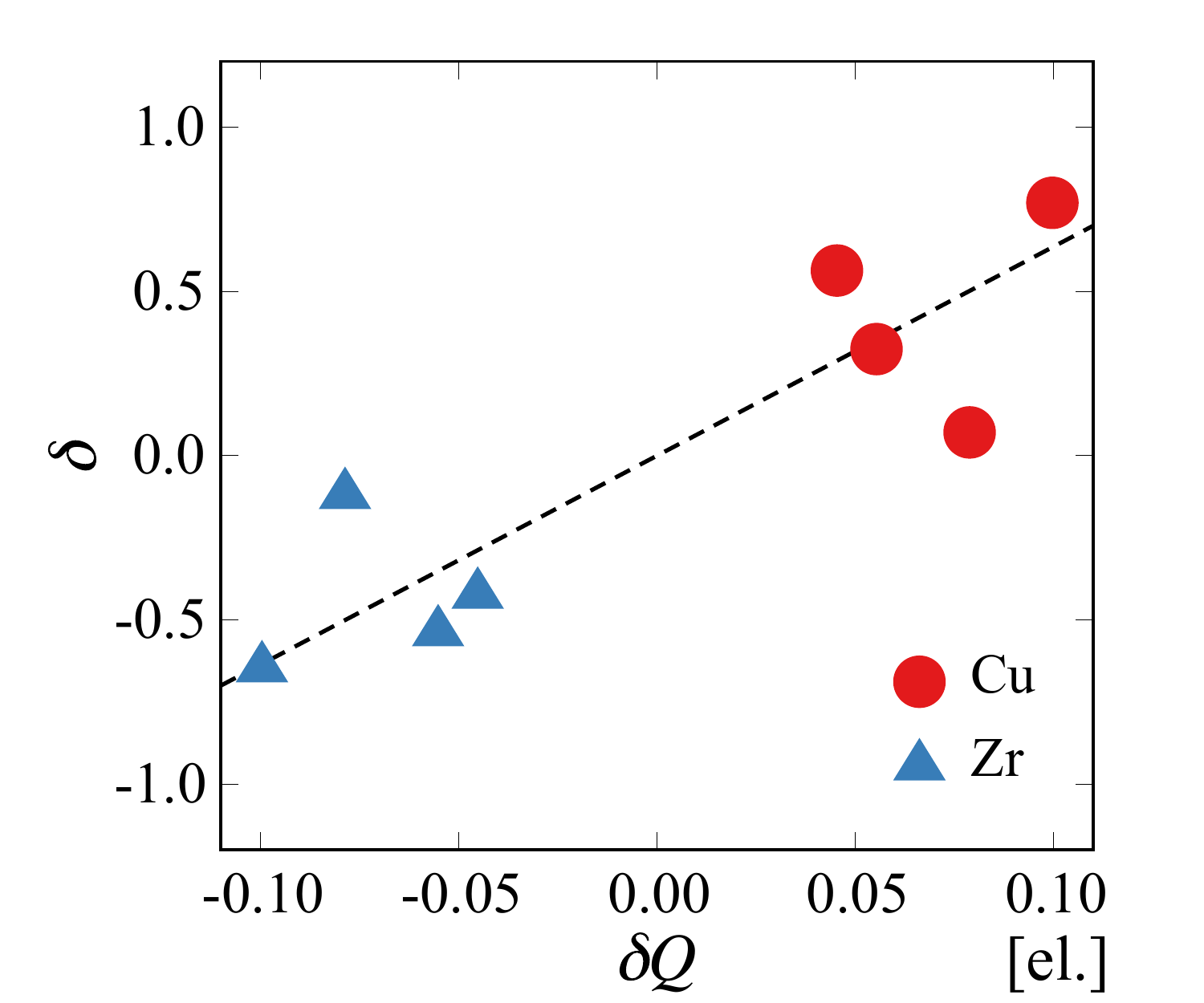}
  \caption{{\em(Color online)} (a) Correlation between the local stress deviation
  ($\delta=(\sigma_\mathit{relaxed}-\sigma_\mathit{affine})/|\sigma_\mathit{affine}|$) and the change in atomic charge ($\delta Q = Q_\mathit{relaxed} - Q_\mathit{affine}$) for Cu$_{50}$Zr$_{50}$ under elastic deformation. Dashed line is guide for eyes. (b) and (c) show the 2D charge density map for one particular structure; shared plane was chosen for atoms Cu0, Cu1, Cu2 to make a section of the 3D charge density.}
  \label{g-dlsdch}
\end{figure*}

Finally, we show a clear correlation between the local shear stress deviation ($\delta=(\sigma_{relaxed}-\sigma_{affine})/|\sigma_{affine}|$) and charge difference, $\delta Q$ for small strains in Fig.~\ref{g-dlsdch}~(a). The data for the highest 8\% strain beyond the elastic region was not included in the figure.
From Fig.~\ref{g-dlsdch}~(b,c), it can be seen how the relaxation affects in the shrinking of the charge clouds around the atom denoted 'Cu0'.
We can conclude that, during the relaxation process, Cu gains charge, which positively increases the stress, whereas Zr loses charge, which makes the stress more negative. It is worth mentioning that there is no direct correlation between the atomic-level pressure deviation and $\delta Q$, although the data are not shown here.

\begin{figure*}
  \centering
  \includegraphics[width=0.95\textwidth]{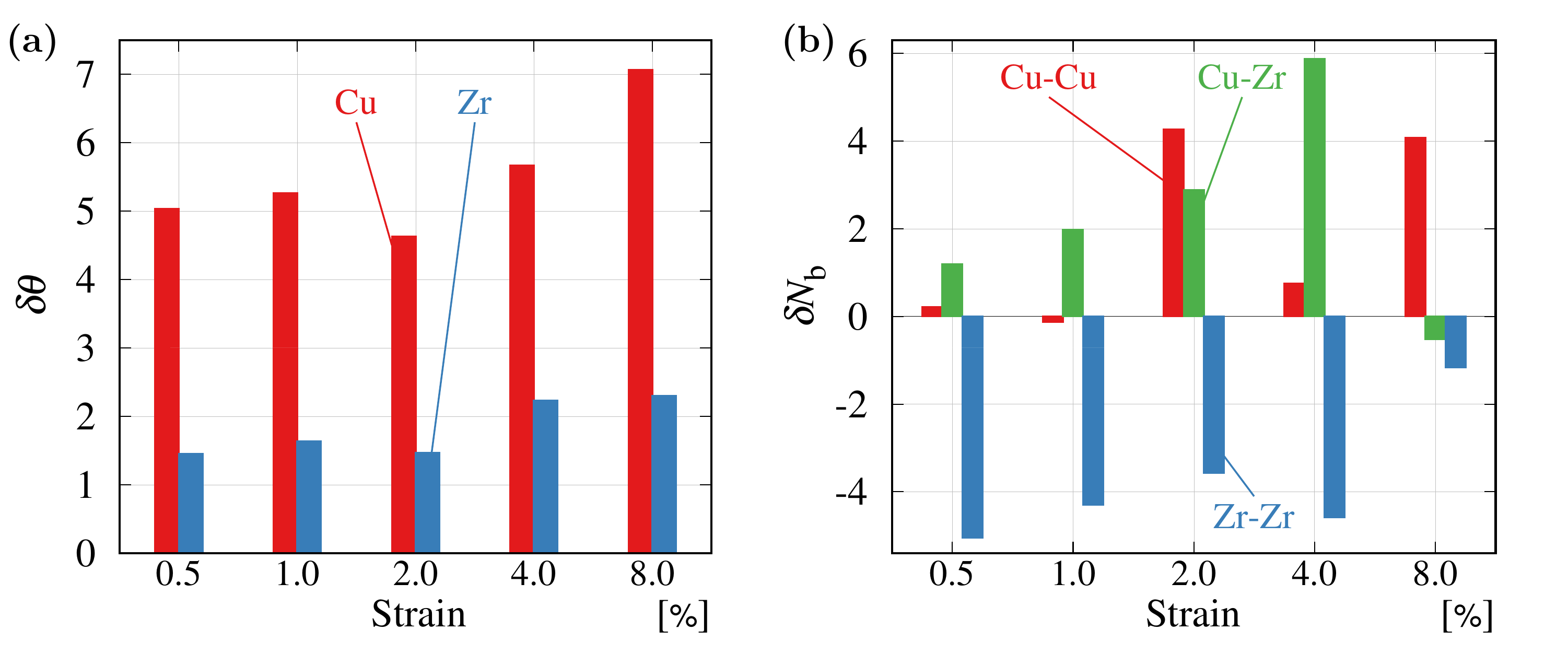}
  \caption{{\em(Color online)} Relaxation induced changes of Cu and Zr subsystems for studied shear strain values. (a) Change in same type angles (see discussion in text). (b) Change in the number of chemical bonds. }
  \label{g-chemNc}
\end{figure*}

% \subsection{Cu/Zr chemical short-range ordering}

To show the difference in the behavior of the Cu and Zr subsystems under strain, let us discuss the 3-atoms angle change induced by the relaxation of atom positions from the affine structure. For each atom, we found closest neighbors of the same type. Next, for that group of atoms, the change of all unique 3-atoms angles was calculated with the targeted atom being the vertex.
All angle change values were averaged and associated with each atom type(see Fig.~\ref{g-chemNc}a). The optimization process for the system under the 8\% shear strain changes angles of the Cu subsystem by the value of 7.06 degrees, while for the Zr subsystem that parameter equals only to 2.24 degrees. The Cu $d$-electron states are full and do not participate in bonding, whereas the Zr $d$-states are only partially filled and form covalent
bonds~\cite{dmowski2017}. The significant difference in the changes in angle induced by relaxation between Cu and Zr is consistent with this difference in the $d$-states occupation.

%% *******
%% ******* ORIGINALY FIG 2
%% *******

Taking only the neighbors of the same type, we can analyze the Cu subsystem and the Zr subsystem separately. However, the interplay between the Cu and Zr subsystems can only be understood if we take into account the types of neighboring atoms (see Supplemental Material for more data). The Cu subsystem is rearranged significantly during the relaxation process, resulting in the stress of an opposite sign, whereas the Zr-Zr angles are affected only slightly. At the same time, the Zr/Cu compositional short-range order (CSRO) is changed by deformation (see Fig.~\ref{g-chemNc}b). The fraction of the Zr-Cu pairs is increased at the expense of the Zr-Zr pairs. The increased CSRO induces more charge transfer, and the lowering of the potential energy, resulting in softening.
An example of actions of bond-breaking and bond-forming is given in Supplemental Material.
However, the CSRO of the system before deformation depends on the condition of the system preparation through the fictive temperature. When we start with the system with a high degree of CSRO, deformation could degrade the CSRO, rather than to enhance it as observed here, resulting in a lesser amount of softening. Such a possibility remains to be studied.

Compelling, we found that the Cu subsystem is affected more strongly by deformation than the Zr subsystem. However, deformation decreases the number of Zr-Zr bonds and increases the Zr-Cu neighbors, thus increasing the charge transfer. Consequently, the stress on Cu is reduced by deformation, leading to the shear softening. Thus, in a Cu$_{50}$Zr$_{50}$ system, not only a geometrical rearrangement of atoms but also a change in a CSRO contributes to the softening of shear modulus.

%%\section{Conclusion}

Summarizing, the effect of elastic deformation on the electronic states in metallic glass is studied under shear strain by means of ab initio calculations. The simulations of Cu$_{50}$Zr$_{50}$ glassy alloy under shear strain show that atoms undergo non-affine deformation even in the elastic regime, accompanied by substantial charge transfer. In particular the Cu subsystem is severely rearranged under the shear strain, surprisingly resulting in the shear stress of the opposite sign on Cu atoms. On the other hand the Zr subsystem deforms in a nearly affine manner. Deformation produces increased Zr/Cu CSRO, decreasing Zr-Zr bonds, which leads to increased charge transfer. This work demonstrates that the change in the CSRO contributes to the softening of the shear modulus through increased charge transfer, along with the geometrical transformation. The evidence from our results points toward the need of considering the effect of the deformation on the CSRO in metallic glasses.

\section{Acknowledgement}

YS is grateful to M. Kohyama and T. Ozaki for discussion and comments. This work was partly supported by Grant-in-Aid for Scientific Research in Innovative Areas to YS (26109705 and 19H05177). TE was supported by the U.S. Department of Energy, Office of Science, Basic Energy Sciences, Materials Science and Engineering Division. TI was supported by JPSJ KAKENHI Grant Number JP19K03771.

\bibliography{references}

\begin{thebibliography}{10}
\expandafter\ifx\csname url\endcsname\relax
  \def\url#1{\texttt{#1}}\fi
\expandafter\ifx\csname urlprefix\endcsname\relax\def\urlprefix{URL }\fi
\expandafter\ifx\csname href\endcsname\relax
  \def\href#1#2{#2} \def\path#1{#1}\fi

\bibitem{inoue1990}
A.~Inoue, T.~Zhang, T.~Masumoto,
  \href{https://www.jstage.jst.go.jp/article/matertrans1989/31/5/31_5_425/_article}{Production
  of amorphous cylinder and sheet of {La$_{55}$Al$_{25}$Ni$_{20}$} alloy by a
  metallic mold casting method}, Materials Transactions, JIM 31~(5) (1990)
  425--428.
\newblock \href {https://doi.org/10.2320/matertrans1989.31.425}
  {\path{doi:10.2320/matertrans1989.31.425}}.
\newline\urlprefix\url{https://www.jstage.jst.go.jp/article/matertrans1989/31/5/31_5_425/_article}

\bibitem{inoue2011}
A.~Inoue, A.~Takeuchi,
  \href{http://www.sciencedirect.com/science/article/pii/S1359645410007822}{Recent
  development and application products of bulk glassy alloys}, Acta Materialia
  59~(6) (2011) 2243 -- 2267.
\newblock \href {https://doi.org/10.1016/j.actamat.2010.11.027}
  {\path{doi:10.1016/j.actamat.2010.11.027}}.
\newline\urlprefix\url{http://www.sciencedirect.com/science/article/pii/S1359645410007822}

\bibitem{johnson1999}
W.~Johnson,
  \href{https://www.cambridge.org/core/journals/mrs-bulletin/article/bulk-glassforming-metallic-alloys-science-and-technology/66E72C89132433493E535F8B817C133D}{Bulk
  glass-forming metallic alloys: Science and technology}, MRS Bulletin 24~(10)
  (1999) 42 -- 56.
\newblock \href {https://doi.org/10.1557/S0883769400053252}
  {\path{doi:10.1557/S0883769400053252}}.
\newline\urlprefix\url{https://www.cambridge.org/core/journals/mrs-bulletin/article/bulk-glassforming-metallic-alloys-science-and-technology/66E72C89132433493E535F8B817C133D}

\bibitem{kumar2011}
G.~Kumar, A.~Desai, J.~Schroers,
  \href{https://onlinelibrary.wiley.com/doi/abs/10.1002/adma.201002148}{Bulk
  metallic glass: The smaller the better}, Advanced Materials 23~(4) (2010)
  461--476.
\newblock \href {https://doi.org/10.1002/adma.201002148}
  {\path{doi:10.1002/adma.201002148}}.
\newline\urlprefix\url{https://onlinelibrary.wiley.com/doi/abs/10.1002/adma.201002148}

\bibitem{davis1976}
L.~Davis, Mechanics of metallic glasses, Rapidly Quenched Metals. Massachusetts
  Institute of Technology, Cambridge. 1976, 369-391 (1976).

\bibitem{weaire1971}
D.~Weaire, M.~Ashby, J.~Logan, M.~Weins,
  \href{http://www.sciencedirect.com/science/article/pii/0001616071901349}{On
  the use of pair potentials to calculate the properties of amorphous metals},
  Acta Metallurgica 19~(8) (1971) 779 -- 788.
\newblock \href {https://doi.org/10.1016/0001-6160(71)90134-9}
  {\path{doi:10.1016/0001-6160(71)90134-9}}.
\newline\urlprefix\url{http://www.sciencedirect.com/science/article/pii/0001616071901349}

\bibitem{suzuki1985}
Y.~Suzuki, T.~Egami,
  \href{http://www.sciencedirect.com/science/article/pii/002230938590242X}{Shear
  deformation of glassy metals: Breakdown of cauchy relationship and
  anelasticity}, Journal of Non-Crystalline Solids 75~(1) (1985) 361 -- 366,
  proceedings of the international conference on the theory of the structures
  of non-crystalline solids.
\newblock \href {https://doi.org/10.1016/0022-3093(85)90242-X}
  {\path{doi:10.1016/0022-3093(85)90242-X}}.
\newline\urlprefix\url{http://www.sciencedirect.com/science/article/pii/002230938590242X}

\bibitem{dmowski2010}
W.~Dmowski, T.~Iwashita, C.-P. Chuang, J.~Almer, T.~Egami,
  \href{https://link.aps.org/doi/10.1103/PhysRevLett.105.205502}{Elastic
  heterogeneity in metallic glasses}, Phys. Rev. Lett. 105 (2010) 205502.
\newblock \href {https://doi.org/10.1103/PhysRevLett.105.205502}
  {\path{doi:10.1103/PhysRevLett.105.205502}}.
\newline\urlprefix\url{https://link.aps.org/doi/10.1103/PhysRevLett.105.205502}

\bibitem{egamiReview2013}
T.~Egami, T.~Iwashita, W.~Dmowski,
  \href{http://www.mdpi.com/2075-4701/3/1/77}{Mechanical properties of metallic
  glasses}, Metals 3~(1) (2013) 77--113.
\newblock \href {https://doi.org/10.3390/met3010077}
  {\path{doi:10.3390/met3010077}}.
\newline\urlprefix\url{http://www.mdpi.com/2075-4701/3/1/77}

\bibitem{sawPRL2016}
S.~Saw, P.~Harrowell,
  \href{https://link.aps.org/doi/10.1103/PhysRevLett.116.137801}{Rigidity in
  condensed matter and its origin in configurational constraint}, Phys. Rev.
  Lett. 116 (2016) 137801.
\newblock \href {https://doi.org/10.1103/PhysRevLett.116.137801}
  {\path{doi:10.1103/PhysRevLett.116.137801}}.
\newline\urlprefix\url{https://link.aps.org/doi/10.1103/PhysRevLett.116.137801}

\bibitem{yuJPCL2016}
H.-B. Yu, R.~Richert, K.~Samwer,
  \href{https://doi.org/10.1021/acs.jpclett.6b01738}{Correlation between
  viscoelastic moduli and atomic rearrangements in metallic glasses}, The
  Journal of Physical Chemistry Letters 7~(19) (2016) 3747--3751.
\newblock \href {https://doi.org/10.1021/acs.jpclett.6b01738}
  {\path{doi:10.1021/acs.jpclett.6b01738}}.
\newline\urlprefix\url{https://doi.org/10.1021/acs.jpclett.6b01738}

\bibitem{argon1979_STZ1}
A.~Argon,
  \href{http://www.sciencedirect.com/science/article/pii/0001616079900555}{Plastic
  deformation in metallic glasses}, Acta Metallurgica 27~(1) (1979) 47 -- 58.
\newblock \href {https://doi.org/10.1016/0001-6160(79)90055-5}
  {\path{doi:10.1016/0001-6160(79)90055-5}}.
\newline\urlprefix\url{http://www.sciencedirect.com/science/article/pii/0001616079900555}

\bibitem{argon1979_STZ2}
A.~Argon, H.~Kuo,
  \href{http://www.sciencedirect.com/science/article/pii/0025541679901745}{Plastic
  flow in a disordered bubble raft (an analog of a metallic glass)}, Materials
  Science and Engineering 39~(1) (1979) 101 -- 109.
\newblock \href {https://doi.org/10.1016/0025-5416(79)90174-5}
  {\path{doi:10.1016/0025-5416(79)90174-5}}.
\newline\urlprefix\url{http://www.sciencedirect.com/science/article/pii/0025541679901745}

\bibitem{langerSTZ2004}
J.~S. Langer,
  \href{https://link.aps.org/doi/10.1103/PhysRevE.70.041502}{Dynamics of
  shear-transformation zones in amorphous plasticity: Formulation in terms of
  an effective disorder temperature}, Phys. Rev. E 70 (2004) 041502.
\newblock \href {https://doi.org/10.1103/PhysRevE.70.041502}
  {\path{doi:10.1103/PhysRevE.70.041502}}.
\newline\urlprefix\url{https://link.aps.org/doi/10.1103/PhysRevE.70.041502}

\bibitem{iwashitaNatComm_Ref18}
J.~D. Ju, D.~Jang, A.~Nwankpa, M.~Atzmon,
  \href{https://doi.org/10.1063/1.3552300}{An atomically quantized hierarchy of
  shear transformation zones in a metallic glass}, Journal of Applied Physics
  109~(5) (2011) 053522.
\newblock \href {https://doi.org/10.1063/1.3552300}
  {\path{doi:10.1063/1.3552300}}.
\newline\urlprefix\url{https://doi.org/10.1063/1.3552300}

\bibitem{iwashitaNatComm_Ref19}
I.-C. Choi, Y.~Zhao, B.-G. Yoo, Y.-J. Kim, J.-Y. Suh, U.~Ramamurty, J.~il~Jang,
  \href{http://www.sciencedirect.com/science/article/pii/S1359646212001340}{Estimation
  of the shear transformation zone size in a bulk metallic glass through
  statistical analysis of the first pop-in stresses during spherical
  nanoindentation}, Scripta Materialia 66~(11) (2012) 923 -- 926, viewpoint set
  no. 49: Strengthening effect of nano-scale twins.
\newblock \href {https://doi.org/10.1016/j.scriptamat.2012.02.032}
  {\path{doi:10.1016/j.scriptamat.2012.02.032}}.
\newline\urlprefix\url{http://www.sciencedirect.com/science/article/pii/S1359646212001340}

\bibitem{iwashitaNatComm_Ref20}
C.~A. Schuh, A.~C. Lund, T.~Nieh,
  \href{http://www.sciencedirect.com/science/article/pii/S1359645404005464}{New
  regime of homogeneous flow in the deformation map of metallic glasses:
  elevated temperature nanoindentation experiments and mechanistic modeling},
  Acta Materialia 52~(20) (2004) 5879 -- 5891.
\newblock \href {https://doi.org/10.1016/j.actamat.2004.09.005}
  {\path{doi:10.1016/j.actamat.2004.09.005}}.
\newline\urlprefix\url{http://www.sciencedirect.com/science/article/pii/S1359645404005464}

\bibitem{iwashitaNatComm_Ref21}
D.~Pan, A.~Inoue, T.~Sakurai, M.~W. Chen,
  \href{http://www.pnas.org/content/105/39/14769}{Experimental characterization
  of shear transformation zones for plastic flow of bulk metallic glasses},
  Proceedings of the National Academy of Sciences 105~(39) (2008) 14769--14772.
\newblock \href {https://doi.org/10.1073/pnas.0806051105}
  {\path{doi:10.1073/pnas.0806051105}}.
\newline\urlprefix\url{http://www.pnas.org/content/105/39/14769}

\bibitem{kohnNobelLecture1999}
W.~Kohn, \href{https://link.aps.org/doi/10.1103/RevModPhys.71.1253}{Nobel
  lecture: Electronic structure of matter\char22{}wave functions and density
  functionals}, Rev. Mod. Phys. 71 (1999) 1253--1266.
\newblock \href {https://doi.org/10.1103/RevModPhys.71.1253}
  {\path{doi:10.1103/RevModPhys.71.1253}}.
\newline\urlprefix\url{https://link.aps.org/doi/10.1103/RevModPhys.71.1253}

\bibitem{fanIwashita2015}
Y.~Fan, T.~Iwashita, T.~Egami,
  \href{https://link.aps.org/doi/10.1103/PhysRevLett.115.045501}{Crossover from
  localized to cascade relaxations in metallic glasses}, Phys. Rev. Lett. 115
  (2015) 045501.
\newblock \href {https://doi.org/10.1103/PhysRevLett.115.045501}
  {\path{doi:10.1103/PhysRevLett.115.045501}}.
\newline\urlprefix\url{https://link.aps.org/doi/10.1103/PhysRevLett.115.045501}

\bibitem{egamiConn1980}
T.~Egami, K.~Maeda, V.~Vitek,
  \href{https://doi.org/10.1080/01418618008243894}{Structural defects in
  amorphous solids a computer simulation study}, Philosophical Magazine A
  41~(6) (1980) 883--901.
\newblock \href {https://doi.org/10.1080/01418618008243894}
  {\path{doi:10.1080/01418618008243894}}.
\newline\urlprefix\url{https://doi.org/10.1080/01418618008243894}

\bibitem{egamiConn2014}
T.~Egami,
  \href{https://www.worldscientific.com/doi/abs/10.1142/S0217984914300063}{Elementary
  excitation and energy landscape in simple liquids}, Modern Physics Letters B
  28 (2014) 1430006 (19pp).
\newblock \href {https://doi.org/10.1142/S0217984914300063}
  {\path{doi:10.1142/S0217984914300063}}.
\newline\urlprefix\url{https://www.worldscientific.com/doi/abs/10.1142/S0217984914300063}

\bibitem{egamiConn2017}
T.~Egami, Y.~Fan, T.~Iwashita,
  \href{https://doi.org/10.1007/978-3-319-45612-6_10}{Mechanical Deformation in
  Metallic Liquids and Glasses: From Atomic Bond-Breaking to Avalanches},
  Springer International Publishing, Cham, 2017, pp. 199--225.
\newblock \href {https://doi.org/10.1007/978-3-319-45612-6_10}
  {\path{doi:10.1007/978-3-319-45612-6_10}}.
\newline\urlprefix\url{https://doi.org/10.1007/978-3-319-45612-6_10}

\bibitem{nicholsonFPStress2013}
D.~M. Nicholson, M.~Ojha, T.~Egami,
  \href{http://stacks.iop.org/0953-8984/25/i=43/a=435505}{First-principles
  local stress in crystalline and amorphous metals}, Journal of Physics:
  Condensed Matter 25~(43) (2013) 435505.
\newblock \href {https://doi.org/10.1088/0953-8984/25/43/435505}
  {\path{doi:10.1088/0953-8984/25/43/435505}}.
\newline\urlprefix\url{http://stacks.iop.org/0953-8984/25/i=43/a=435505}

\bibitem{matternCuZr2008_1}
N.~Mattern, A.~Schops, U.~Kuhn, J.~Acker, O.~Khvostikova, J.~Eckert,
  \href{https://www.infona.pl/resource/bwmeta1.element.elsevier-a012e0d7-677c-3b84-883e-b5a431eb7b33}{Structural
  behavior of cu x zr 100 x metallic glass \( x = 35-70\)}, Journal of
  Non-crystalline Solids 354 (2008) 1054--1060.
\newblock \href {https://doi.org/10.1016/j.jnoncrysol.2007.08.035}
  {\path{doi:10.1016/j.jnoncrysol.2007.08.035}}.
\newline\urlprefix\url{https://www.infona.pl/resource/bwmeta1.element.elsevier-a012e0d7-677c-3b84-883e-b5a431eb7b33}

\bibitem{matternCuZr2008}
N.~{Mattern}, A.~{Sch{\"o}ps}, U.~{K{\"u}hn}, J.~{Acker}, O.~{Khvostikova},
  J.~{Eckert},
  \href{http://adsabs.harvard.edu/abs/2008JNCS..354.1054M}{{Structural behavior
  of CuxZr100-x metallic glass (x=35-70)}}, Journal of Non Crystalline Solids
  354 (2008) 1054--1060.
\newblock \href {https://doi.org/10.1016/j.jnoncrysol.2007.08.035}
  {\path{doi:10.1016/j.jnoncrysol.2007.08.035}}.
\newline\urlprefix\url{http://adsabs.harvard.edu/abs/2008JNCS..354.1054M}

\bibitem{vasp}
The vasp web-site, https://www.vasp.at.

\bibitem{vasp_mp}
M.Methfessel, A.T.Paxton,
  \href{https://journals.aps.org/prb/abstract/10.1103/PhysRevB.40.3616}{High-precision
  sampling for brillouin-zone integration in metals}, Physical Review B 40
  (1989) 3616.
\newblock \href {https://doi.org/10.1103/PhysRevB.40.3616}
  {\path{doi:10.1103/PhysRevB.40.3616}}.
\newline\urlprefix\url{https://journals.aps.org/prb/abstract/10.1103/PhysRevB.40.3616}

\bibitem{shiiharaLS2010}
Y.~Shiihara, M.~Kohyama, S.~Ishibashi,
  \href{https://link.aps.org/doi/10.1103/PhysRevB.81.075441}{Ab initio local
  stress and its application to al (111) surfaces}, Phys. Rev. B 81 (2010)
  075441.
\newblock \href {https://doi.org/10.1103/PhysRevB.81.075441}
  {\path{doi:10.1103/PhysRevB.81.075441}}.
\newline\urlprefix\url{https://link.aps.org/doi/10.1103/PhysRevB.81.075441}

\bibitem{openMX}
The openmx web-site, http://www.openmx-square.org/.

\bibitem{nakamura2014}
A.~Nakamura, Y.~Kamimura, K.~Edagawa, S.~Takeuchi,
  \href{http://www.sciencedirect.com/science/article/pii/S0921509314008648}{Elastic
  and plastic characteristics of a model cuzr amorphous alloy}, Materials
  Science and Engineering: A 614 (2014) 16 -- 26.
\newblock \href {https://doi.org/10.1016/j.msea.2014.07.010}
  {\path{doi:10.1016/j.msea.2014.07.010}}.
\newline\urlprefix\url{http://www.sciencedirect.com/science/article/pii/S0921509314008648}

\bibitem{dmowskiActaMater2010}
W.~Dmowski, Y.~Yokoyama, A.~Chuang, Y.~Ren, M.~Umemoto, K.~Tsuchiya, A.~Inoue,
  T.~Egami, \href{http://dx.doi.org/10.1016/j.actamat.2009.09.021}{Structural
  rejuvenation in a bulk metallic glass induced by severe plastic deformation}
  58~(2) (2010) 429--438.
\newblock \href {https://doi.org/DOI:101016/jactamat200909021}
  {\path{doi:DOI:101016/jactamat200909021}}.
\newline\urlprefix\url{http://dx.doi.org/10.1016/j.actamat.2009.09.021}

\bibitem{dmowski2017}
W.~Dmowski, S.~Gierlotka, Z.~Wang, Y.~Yokoyama, B.~Palosz, T.~Egami,
  \href{https://doi.org/10.1038/s41598-017-06890-w}{Pressure induced
  liquid-to-liquid transition in zr-based supercooled melts and pressure
  quenched glasses}, Scientific Reports 7~(1) (2017) 6564.
\newblock \href {https://doi.org/10.1038/s41598-017-06890-w}
  {\path{doi:10.1038/s41598-017-06890-w}}.
\newline\urlprefix\url{https://doi.org/10.1038/s41598-017-06890-w}

\end{thebibliography}

\end{document}